\documentclass[preprint2]{aastex}

\slugcomment{}

\shorttitle{Optical and Near-IR study of nova V2676 Oph 2012}
\shortauthors{Raj et al.}

\begin{document}


\title{Optical and Near-IR study of nova V2676 Oph 2012}


\author{A. Raj \altaffilmark{} 
\affil{Korea Astronomy and Space Science Institute, Daejeon, 34055, Korea.}
\affil{Indian Institute of Astrophysics, II Block Koramangala, Bangalore 560 034, India.}
R. K. Das\altaffilmark{}
\affil{Department of Astrophysics \& Cosmology, S N Bose National Centre for Basic Sciences, Salt Lake, Kolkata 700106, India.}
F. M. Walter\altaffilmark{} 
\affil{Department of Physics and Astronomy, Stony Brook University, Stony Brook, NY 11794-3800, USA.}
\email{ashish.raj@iiap.res.in}
}
\begin{abstract}
We present optical spectrophotometric and near-infrared (NIR) photometric observations of the nova V2676 Oph covering the period from 2012 March 29 through 2015 May 8. 
The optical spectra and photometry of the nova have been taken from SMARTS and Asiago; the near-infrared photometry was obtained from SMARTS and Mt. Abu.  
The spectra are dominated by strong H {\sc i} lines from the Balmer series, Fe\,{\sc ii}, N\,{\sc i} and [O\,{\sc i}] lines in the initial days, typical of an Fe {\sc ii} type nova.
The measured FWHM for the H$\beta$ and H$\alpha$ lines was 800-1200 km s$^{-1}$. There was pronounced dust formation starting 90 days after the outburst. The $J-K$ color was the largest among recent 
dust forming novae. 
\end{abstract}


\section{Introduction}

The classical novae are interacting binary star systems containing a Roche-lobe filling secondary, on or near the main sequence, which is losing
hydrogen-rich material through the inner Lagrangian point to the degenerate white dwarf primary. The mass transfer results in the formation of an accretion disc around 
the white dwarf. 
The runway thermonuclear reactions on the white dwarf surface give rise to the thermonuclear outburst, the sudden brightening seen 
in these systems.

Nova V2676 Oph (Nova Oph 2012) was discovered on 2012 March 25.789 UT (which we define as t=0) by Hideo Nishimura on three 13s unfiltered CCD frames with limiting magnitude 
13.5 at $V$ = 12.1 (Nishimura et al. 2012). A low resolution optical spectrum obtained on March 27.74 UT with the 1.3m Arakai telescope at Koyama Astronomical Observatory 
(Arai \& Isogai 2012) showed H$\alpha$, H$\beta$, and O {\sc i} lines having prominent P-Cygni profiles. The FWHM of the emission component of H$\alpha$ was about
600 km s$^{-1}$. They suggested that the object is an Fe {\sc ii}-type classical nova.
Another low-resolution spectrum taken by Imamura (2012) at similar time on March 27.836 UT showed prominent emission lines of H$\alpha$, 
H$\beta$ and Fe {\sc ii} 
which confirmed that the nova was of Fe {\sc ii} class. 

The near-IR observations taken between March 28-30 UT also showed that the spectra are typically
of a Fe {\sc ii} class nova having prominent H {\sc i} emission lines of Pa$\beta$, Pa$\gamma$ and Br$\gamma$, Fe {\sc ii} and other
Brackett-series lines (Rudy et al. 2012a). The other prominent features seen were C {\sc i}, O {\sc i}, N {\sc i} and 
Ca {\sc ii} infrared triplet. 
Rudy et al. (2012b) reported the fundamental, first and second overtone bands of 
CO in emission on 2012 May 1 and 2 and suggested a strong possibility of dust formation in the nova ejecta.

In this paper we present optical and NIR observations of V2676 Oph. 
The outline of the paper is as follows: Section 2 describes the observations and data analysis techniques. 
The results obtained from these observations are discussed in Section 3 and the summary is given in Section 4.

\section{Observations}

\subsection{Near-infrared observations}

Near-IR observations were obtained using the 1.2m telescope of Mt.Abu Infrared Observatory from 2012 March 29 to 2012 June 17 and 
SMARTS/CTIO 1.3m telescope with the Andicam dual channel photometer (see Walter et al. 2012 for a description of the of the 
instrument and the data reductions). 
The log of the photometric observations from Mt. Abu is given in Table 1 and the SMARTS photometry is available on SMARTS atlas. 
Photometry in the $JHK$ bands was done under clear sky conditions using 
a near-infrared Imager/Spectrometer with a 256$\times$256 HgCdTe NICMOS3 array in the imaging mode. Several frames, at 4 dithered positions, 
offset by $\sim$ 30 arcsec were obtained in all the bands. The sky frames, which are subtracted from the nova frames, were generated by median-combining 
the dithered frames (see Raj et al. 2013 for more details). The standard star SAO 185406 (spectral type - B9.5/A0V) having $JHK$ magnitudes 6.47, 6.53 and 6.53 respectively, 
was used for photometric calibration. The data are reduced and analyzed using the $IRAF$ package.

\subsection{Optical observations}

Optical spectra were obtained with the Asiago 1.22m telescope + B\&C spectrograph, 2 arcsec slit width and oriented along North-South. 
The calibration in absolute fluxes for each nova spectrum was done by observations of several spectrophotometric standards observed on 
same night around similar airmasses. More detailed description of ANS (Asiago Novae and Symbiotic stars) Collaboration instruments, operation modes and results on the monitoring of 
novae can be found in Munari et al. (2012). 

Further low dispersion spectra and photometry were obtained using the SMARTS Facilities\footnote{http://www.astro.sunysb.edu/fwalter/SMARTS/NovaAtlas}. The R-C grating spectrograph, 
the data reduction techniques, and the
observing modes are described by Walter et al. (2012). Basically, mode 13/I is sensitive to the entire optical band, from
3200 \AA\ through 9500 \AA; mode 26/Ia is sensitive from 3650 \AA\ through 5400 \AA; mode 47/I covers 5650 through 6900 \AA, and mode 47/IIb covers 4060 \AA\ through
4720 \AA. We obtained 19 spectra on an irregular cadence and with various sky conditions from 2012 April 5 through 2012 June 24.
The target was observed using the COSMOS\footnote{http://www.ctio.noao.edu/noao/content/cosmos} long slit spectrograph at CTIO on 2015 May 8, some
3 years after outburst. 
We used the r2k disperser in combination with the 3pxR slit and the OG570 filter to obtain wavelength coverage from 6082 \AA\ through 10253 \AA, with
a reciprocal dispersion of 1 \AA/pixel, and a resolution of about 3 \AA. Data reductions are similar to those used for the SMARTS spectra.
The log of the spectroscopic observations is given in Table 2.

\section{Results}

\subsection{General characteristics of $V$ and $JHK$ band light curves}

The $V$ and $JHK$  band light curves are made using the data from American Association of Variable Star Observers (AAVSO), SMARTS/CTIO 1.3m 
(Walter et al. 2012), Asiago 1.22m and Mt. Abu 1.2m telescope Facilities (see Fig. 1). 
The $V$ band light curve shows fluctuations of $\sim$ 0.7 magnitude around m$_v$ = 11.5 
up to 70 days from outburst before beginning a slow decline. 
A sudden drop of more than 5 mag commencing after 90 days clearly indicates the dust formation in the nova ejecta 
(see Fig. 1, lower panel). The nova faded to $>$ 23~mag
between 110-210 days from the outburst, suggesting that a large amount of dust
was formed in the nova ejecta. After day 215 it recovered to $V\sim$~18. Since day 400 it has faded slowly, to $B,V,R,I$ magnitudes 
of about 21, 18.7, 18.3, and 19.2, respectively, in 2016 February.

The near-infrared $JHK$ light curves are made using the data from Mt. Abu observations (see Table 1) and SMARTS/CTIO 1.3m telescope facility  (Walter et al. 2012). 
The NIR light curves are complex. The initial trends are downward in $J$ and more-or-less flat in
$H$ and $K$, with superposed fluctuations similar to those seen in the optical (see Fig.~1). After about day 70 
the $K$ band brightness increased by over 2 mag, accompanied by a smaller brightening in $H$,
while $J$ continued to fade.  

The K band brightness reached a peak value of $\sim$ 5.2 on day 93, the $J-K$ and $H-K$ colors were about 3.1 (marked in Fig. 1) 
and 1.5, respectively.
These NIR color excess indicate that dust has been formed in the nova ejecta. 
Thereafter, the $JHK$ band light curves show a steep steady decline 
from 100-190 days from the outburst, lagging the dust dip seen in the optical. 
After day 190, the NIR brightness started recovering to brighter levels. The deep NIR dip suggests that the dust emission is 
optically thick 
even out to 2.2$\mu$m. The dip is both shallower and narrower than the optical dip, consistent
 with dust opacity.
After recovering, the NIR fluxes faded at a rate of about 0.015 mag/day until the target
became too faint to follow with Andicam after 2014 August 14 (about day 700). At that time
the J-K color had decreased to about 3 mag, which is consistent with a 1500-2000 K blackbody dust shell. Varricatt et al. (2013) 
reported NIR photometry between 2012 Sept 6 and 2013 Apr 9 which confirms the presence of the dust shell.

\subsection{The reddening and distance of V2676 Oph}

From the optical spectra presented in this paper, we estimate the reddening A$_V$ = 2.9 $\pm$ 0.1 for 
the V2676 Oph using Balmer decrement method assuming the electron density $\sim$ 10$^6$ cm$^{-3}$ and temperature $\sim$ 5000K 
(see section 3.5 for more details). However, it should be noted that the hydrogen-recombination lines are usually not described by the Case B 
approximation in novae and consequently the uncertainty in E(B-V) from this method is likely to be 
underestimated. The value for A$_V$ derived earlier is comparable with other  
estimated values for reddening e.g. A$_V$ = 2.39 $\pm$ 0.12 (Nagashima et al. 2015) and A$_V$ = 2.67 $\pm$ 0.16 (Kawakita et
al. 2016) for R = 3.1 in the case of V2676 Oph. Alternatively, we can find the value of extinction from reddening E(B - V) $\sim$ 0.93 towards the nova direction taken 
from Schlafly \& Finkbeiner (2011) which was estimated using the colors of stars with spectra in the Sloan Digital Sky Survey. 
This gives interstellar extinction A$_V$ $\sim$ 2.89 for R = 3.1 towards
the nova direction. Though the value of A$_V$ derived from our spectra is close to the
Schlafly \& Finkbeiner (2011) value, the values for electron density and temperature are specific for the region where [O\,{\sc i}] 
forms and are unlikely applicable to the region containing the H. Hence, for we use the value derived by 
Schlafly \& Finkbeiner (2011) for reddening.

Since the brightness of the nova does not decline smoothly and there is large drop in magnitude due to dust formation, it is 
difficult to measure the characteristic time t$_2$, the time to decline by two magnitude from visual maximum, of the nova, directly
from the light curve. 
Instead, we can use the relation shown by Williams et al. (2013)
to show that, given that the dust starts to condense about day 90, t$_2$ is likely to be between 60 and 80 days. 
This indicates that V2676 Oph is a moderately fast nova. 
Applying maximum magnitude rate of decline (MMRD) relation by Downes \& Duerbeck (2000) and the extinction value as
mentioned above, we estimate the range for M$_{V_{max}}$ between -6.5 to -6.8. Using this range for M$_{V_{max}}$ together with the value 
of the visual maximum V$_{max}$ = 10.6 on April 5, we estimate the range for the distance d to the nova between 6.9 - 7.9 kpc 
and the height z of
the nova to be in the range 637 - 730 pc above the Galactic plane. The large reddening is consistent with the
location near the Galactic center (l,b = 0.26, +5.30) and the large distance.

\subsection{Line identification, evolution and general characteristics of the optical spectra}

The optical spectra from SMARTS and Asiago, presented in Fig. 2 cover the pre-maximum to the early decline phase with one spectrum in the nebular phase. 
We obtained 19 low dispersion spectra using the SMARTS 1.5m/RC spectrograph from  2012 April 5 through 2012 June 24 (days 11 through 91). 
The initial spectrum covered the entire available spectrum at low dispersion. H$\alpha$ was in emission (EW $\sim$ -7.3 \AA ), with a P-Cygni absorption evident 
(EW $\sim$ 0.6 \AA). 
No other lines are seen in emission. The H {\sc i} Balmer series is seen in absorption at least through H-10, and Paschen lines Pa9 - Pa12 also seem to be present.
At this low dispersion absorption lines can be hard to identify unambiguously, but other strong absorption lines include Ca {\sc ii}~K \& H, Na {\sc i} (5890/5896; 6154/6160, 
8191), possible O {\sc i} 7774, and numerous lines that may be Fe {\sc ii} multiplets 27, 28, 37, 38 and 74.
By day 22 the strength of the H$\alpha$ emission had grown to an equivalent width of -63 \AA. At 3.1 \AA\ resolution the line is asymmetric, with a notch
about 350 km s$^{-1}$ blue-ward of the emission peak. Na {\sc i}~D is in emission with a strong P-Cygni absorption feature. The Na {\sc i} 6154/6160 lines are in emission, with
absorption components at -600 km s$^{-1}$. He {\sc i} 6678 is in emission. Strong absorption may be P-Cygni outflows (and weak emission) in Si {\sc ii} 6347 and various Fe {\sc ii} 
multiplet 74 lines.

A low resolution spectrum on day 26 shows many emission lines, including the Balmer lines through H-10, the Fe {\sc ii} multiplet 42, 48, 49 and 55 lines,
and the Ca {\sc ii} infrared triplet.
The equivalent width of H$\alpha$ remained at -63 \AA. Prominent P-Cygni lines are seen at Na {\sc i} D and 6154/6160, O {\sc i} 7002, 7774, 8446, C {\sc i} 7120, N {\sc i} 7450,
Ca {\sc ii}-K \& H are in absorption.
On day 28 the Na {\sc i}~D1 and D2 absorption components are well-defined, at -660 km s$^{-1}$.
The H$\alpha$ equivalent width had increased to -165 \AA; the line profile is asymmetric with
a notch 320 km s$^{-1}$ blue-ward of the peak and a red tail extending to about 1500 km s$^{-1}$.
Our first blue spectrum, on day 34, shows a typical Fe {\sc ii} nova spectrum. H$\beta$ is the strongest line, followed by Ca {\sc ii}-K \& H and the Fe {\sc ii} multiplet
42 lines. The Fe {\sc ii} multiplet 42 lines show narrow P-Cygni absorption at -950 km s$^{-1}$. By day 46 this narrow fast P-Cygni outflow was also visible 
in the Balmer lines (H$\beta$ through H-11).
This absorption component persisted at least through day 67, by which time it had accelerated to a velocity of about -1000 km s$^{-1}$. The first two Asiago
spectra were obtained in the midst of these, on days 45 and 52.
The strong outflow in the Na {\sc i}~D lines which had persisted through day 69 (EW $\sim$ 9 \AA) weakened by about a factor of 10 by day 89. At this time, 
based on a P-Cygni line profile at He {\sc i} 6678, there may be similar emission at He {\sc i} 5876 complicating the line profile. Also by day 89, 
H$\alpha$ and the 6300/6364 \AA [O {\sc i}] lines are developing double-horned line profiles, 
with V$>$R. The H$\alpha$ equivalent width was -400 \AA.
In our last (blue) spectrum, on day 91, or about a week after the final Asiago spectrum,
the P-Cygni absorption components (velocity $\sim$-1100 km s$^{-1}$) of the Balmer lines remain prominent, while those of the Fe {\sc ii} lines are less distinct.
The re-appearance of P-Cygni profiles in the later phase have been seen in other novae e.g. V1186 Sco, V2540 Oph, 
V4745 Sgr, V5113 Sgr, V458 Vul, and V378 Ser (Tanaka et al. 2011). This can be attributed to a re-expansion of the photosphere 
(see Tanaka et al. 2011 for more details).
   
The COSMOS spectrum taken on 2015 May 8 shows that the nova was in the nebular phase. 
In the 6100-10200 \AA\ spectral range the strongest line by far is H$\alpha$+[N {\sc ii}] 6548/6584,
followed by the forbidden lines of [S {\sc iii}] 9531, the four [O {\sc ii}] lines from 7319-7331 \AA, 
and [O {\sc i}] 6300 and [Ar {\sc iii}] 7751. The only permitted lines are 
H$\alpha$ and some He {\sc i} and He {\sc ii} lines. 
Overall the spectrum resembles that of V443 Sct some 2 years after its outburst (Williams, Phillips \& Hamuy, 1994).
The H$\alpha$+[N {\sc ii}] lines appear to be resolved, and can be fit as a sum of 3 Gaussians.
The median FWHM is 623 km s$^{-1}$, while the instrumental resolution is about 100 km s$^{-1}$. 
Radial velocities are measured from the line centroids. The median radial velocity is 6 km s$^{-1}$, with an uncertainties of order
50 km s$^{-1}$, from convolving measurement errors with uncertainties in line rest wavelengths.  
Our observations are consistent with the results of Nagashima et al. (2014, 2015) and Kawakita et al. (2015, 2016). 
A list of the prominnet lines identified and the reddening-corrected emission line fluxes (in erg s$^{-1}$ cm$^{-2}$) for selected epochs is given in 
Tables 3, 4 and 5. The complete tables are available in the online version of the paper.

\begin{figure}
\centering
 \includegraphics[width=2.4 in]{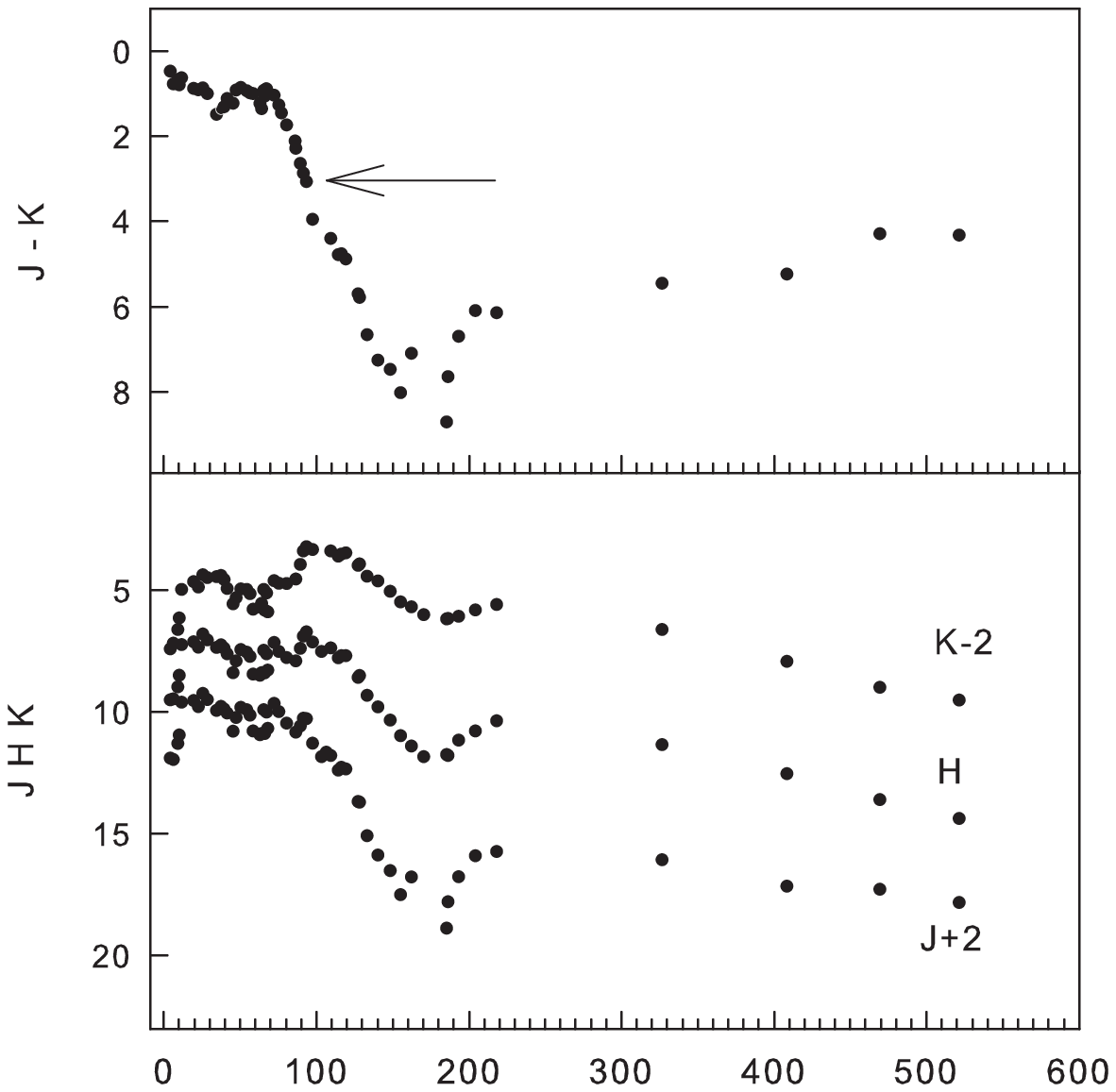}
 \includegraphics[scale=0.31, angle=-90]{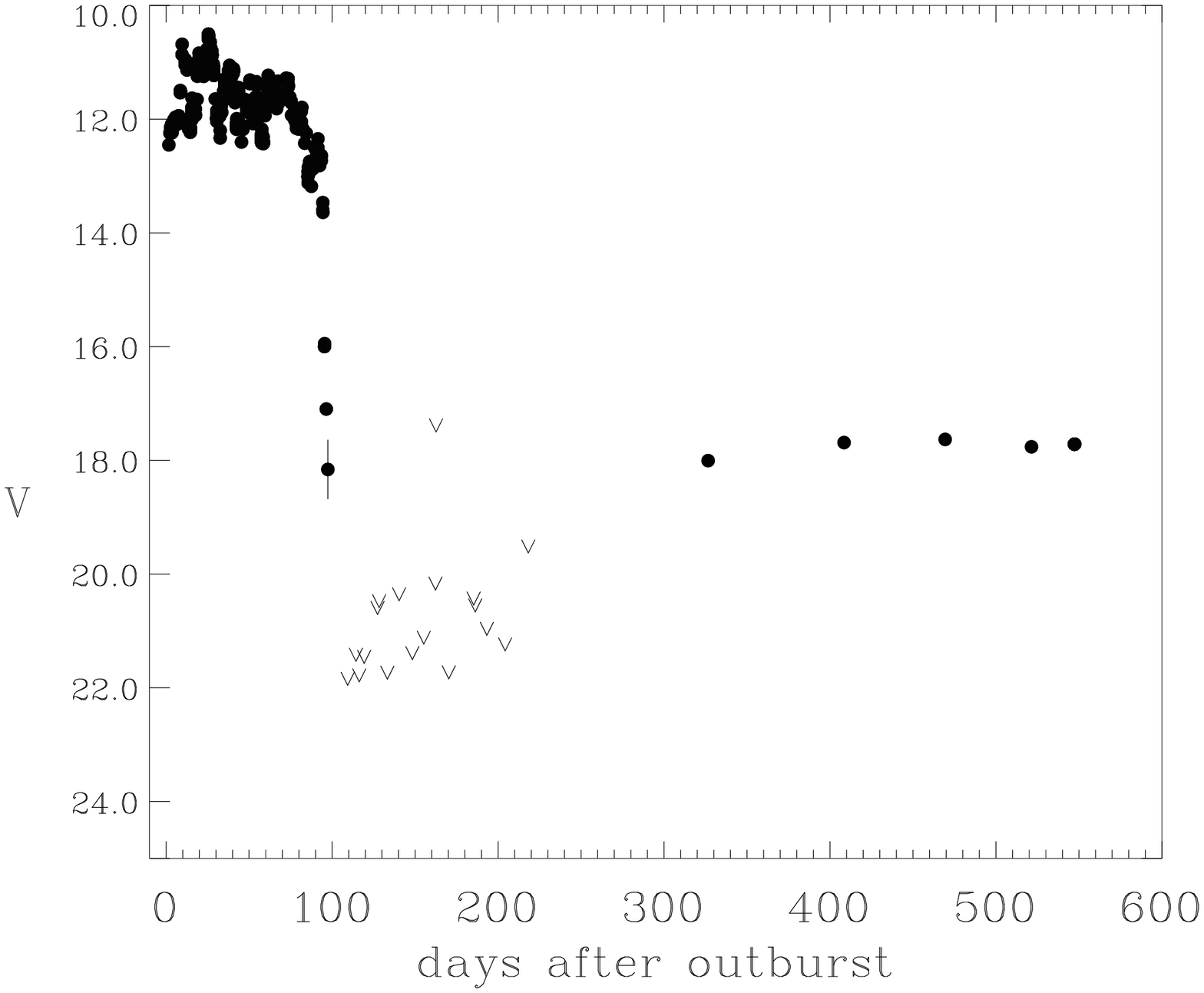}
\caption{The $V$ and $JHK$ band light curves of V2676 Oph based on the data obtained from Mt. Abu, Asiago and SMARTS/CTIO facilities, 
are presented. The sudden fall in the $V$ band light curve (lower panel) after 90 days from the outburst clearly indicates the onset 
of dust formation in the nova ejecta. Upper limits are 1$\sigma$.}
\end{figure}

\subsection{Dust formation and $J - K$ colors}   

A sharp decline in the optical light curve (the dust dip) about 90 days after the outburst and a steady increase in the near-IR magnitudes, especially in the K band clearly
indicate the onset of the dust formation. 
The dust dip in the $BVRI$ band light curves is very similar to the nova V5668 Sgr which, like
V2676~Oph, also showed the CO band in emission 
(Banerjee et al. 2016). 
The presence of CO molecules in early nova spectra are
indicators of low-temperature zones which are conducive to dust formation in the nova ejecta. This is clearly confirmed in the case of V2676 Oph. However, 
only 10 novae have shown the CO bands in emission before the dust formation so far (Banerjee et al. 2016, Raj et al. 2015) and there are so many cases such as 
V1280 Sco (see Das et al. 2008), V5579 Sgr (Raj et al. 2011) where dust formation was reported without any CO emission. 
The reason is not clearly understood; possibly these molecules were present but the strengths were below detection levels or due to 
observational constraints they might have formed and destroyed before the detection.

An increase in flux in the H and K bands coincides with the onset of the dust dip at shorter wavelengths. 
This is likely thermal emission from the dust. The larger excess in K is consistent with expectations that the thermal emission 
will peak in or beyond the K band.
Similar behavior was seen in V5579 Sgr (Raj et al. 2011) and V496 Sct (Raj et al. 2012), which also showed the dust formation. 
The $J-K$ color excess reached a maximum of $\sim$8~mag at 190 days from outburst.
This seems to be the largest $J-K$ value observed for the dust forming novae in recent 
years; ($J-K$) = 2.76 for V5579 Sgr (Raj et al. 2011), 3.79 for V496 Sct (Raj et al. 2012), 4.58 for V5584 Sgr (Raj et al. 2015), 
4.65 for V1280 Sco (Das et al. 2008) and 3.97 for V2615 Oph (Das et al. 2009). 
Thereafter the $J-K$ color decreased, presumably because the
dust thinned due to geometrical dilution. 

The presence of strong C {\sc i} lines and the early dust formation (just after 90 days) in the case of V2676 Oph 
indicate the presence of carbon grains in the dust shell (Clayton \& Wickramasinghe 1976). This is further supported by the detection of C$_2$ and CN molecules in 
V2676 Oph in the early phase (Nagashima et al. 2014). 
  
\begin{table}
\scriptsize
\caption{Log of the $JHK$ photometric observations of nova V2676 Oph from Mt. Abu.}

\begin{tabular}{llcccc}
\hline \\
Date of&     Days after  & & Magnitudes &     \\
observation&    discovery  & &              &     \\
2012 (UT) &         &$J$ &$H$          &$K$ \\

\hline \\

Mar 30  & 04&9.89 $\pm$ 0.05   & 9.50 $\pm$ 0.06  & 9.41 $\pm$ 0.15  \\

Apr 01  &06& 9.95 $\pm$ 0.04  & 9.46 $\pm$ 0.12 & 9.17 $\pm$ 0.15 \\

Apr 03  &08& 9.29 $\pm$ 0.04  & 8.97 $\pm$ 0.09 & 8.61 $\pm$ 0.16  \\

Apr 04   & 09&8.94 $\pm$ 0.04  & 8.49 $\pm$ 0.09 & 8.14 $\pm$ 0.10  \\

May 27   & 62&8.93 $\pm$ 0.07  & 8.49 $\pm$ 0.09 & 7.69 $\pm$ 0.07  \\

May 28  &63 &8.89 $\pm$ 0.07   & 8.38 $\pm$ 0.03 & 7.53 $\pm$ 0.21   \\

May 29   & 64&8.86 $\pm$ 0.03  & 8.46 $\pm$ 0.03 & 8.29 $\pm$ 0.09  \\

May 30  &65& 8.89 $\pm$ 0.04   & 8.39 $\pm$ 0.01 & 7.82 $\pm$ 0.12 \\

May 31   &66& 9.18 $\pm$ 0.08  & 8.70 $\pm$ 0.05 & 8.28 $\pm$ 0.09  \\

June 01  &67&8.67 $\pm$ 0.04   & 8.28 $\pm$ 0.03 & 7.88 $\pm$ 0.12  \\

June 02   &68 &9.10 $\pm$ 0.04  & 8.75 $\pm$ 0.13 & 8.59 $\pm$ 0.03 \\

June 03   &69&8.73 $\pm$ 0.04   & 8.40 $\pm$ 0.07 & 8.42 $\pm$ 0.03   \\

June 07   & 73&8.61 $\pm$ 0.05  & 8.29 $\pm$ 0.03 & 8.30 $\pm$ 0.20  \\

June 09  & 75&8.81 $\pm$ 0.01   & 8.29 $\pm$ 0.03 & 7.35 $\pm$ 0.19   \\

June 18   &84&9.39 $\pm$ 0.06   & 8.53 $\pm$ 0.02 & 7.27 $\pm$ 0.13   \\

\hline
\end{tabular}
\end{table}
  
\subsection{Physical Parameters}

The optical spectra can be helpful for estimating the physical parameters of the nova ejecta with the use of hydrogen and oxygen line fluxes. 
The electron number densities are large in the early phase of nova evolution thus the optical depth $\tau$ which is an important parameter can be used to estimate the elemental abundances. 
Using the formulation of Williams (1994), 
 \begin{displaymath}%
\frac{j_{6300}}{j_{6364}}
=\frac{1-{\rm e}^{-\tau}}{1-{\rm e}^{-\tau/3}}
\end{displaymath}
we estimate the optical depth $\tau$ for the [O {\sc i}] 6300 \AA\ line for the period between 2012 May 10 to 2012 June 16, 
in the range of 1.1-1.7. 
Using the value of $\tau$ we can estimate the electron temperature given by the relation:
\begin{displaymath}%
T_{e}=\frac{11~200}{\log~[43\tau/(1-{\rm e}^{-\tau})\times
F_{\lambda6300}/F_{\lambda5577}]}
\end{displaymath}
we find T$_e$ $\sim$ 4400 K which is consistent with other novae (Ederoclite et al. 2006). 
To estimate the mass of the hydrogen m(H), we
use the following relation from Osterbrock \& Ferland  (2006) which relates the intensity of the H$\beta$ line and the mass of
the hydrogen in the emitting nebula having pure hydrogen as, 
\begin{displaymath}%
m(H)/M{_\odot} = d^2 \times 2.455 \times 10^{-2} \times I(H\beta)/\alpha_{eff} N_e
\end{displaymath}

where $\alpha$ $_{eff}$ is the effective recombination coefficient taken from Storey \& Hummer (1995) and I$(H\beta)$ 
is the flux for H$\beta$ line. We have used two I$(H\beta)$ flux values as 6.33 $\times$ 10$^{-11}$ and 5.19 $\times$ 10$^{-11}$ erg cm$^{-2}$ sec$^{-1}$ for 
May 17, 2012 and June 16, 2012, respectively. 
The [O I] 6300, 6364 and 5577 \AA\ lines are used to set the range for the electron density ($N_e$), 10$^6$ - 10$^7$ cm$^{-3}$.
We do not notice any significant change in T$_e$ and N$_e$ estimated above for both the epochs.
The mass of the hydrogen m(H) is estimated as (2.7 $\times$ 10$^{-7}$ - 3.3 $\times$ 10$^{-6}$)$d^2$M${_\odot}$ where $d$ is the
distance to the nova. 

\section{Summary}

We have presented the optical spectrophotometry and near-IR photometry of nova V2676 Oph which erupted in late-March 2012. 
The optical spectra indicate that the nova belongs to Fe {\sc ii} class. The reddening and distance to the nova are calculated. The nova showed a large amount of 
dust formation 90 days after the outburst.
The physical parameters, optical depth, electron temperature and the mass of hydrogen are estimated in V2676 Oph. 

\begin{figure*}
\centering
 \includegraphics [height=4.8in, width=6.5in,]{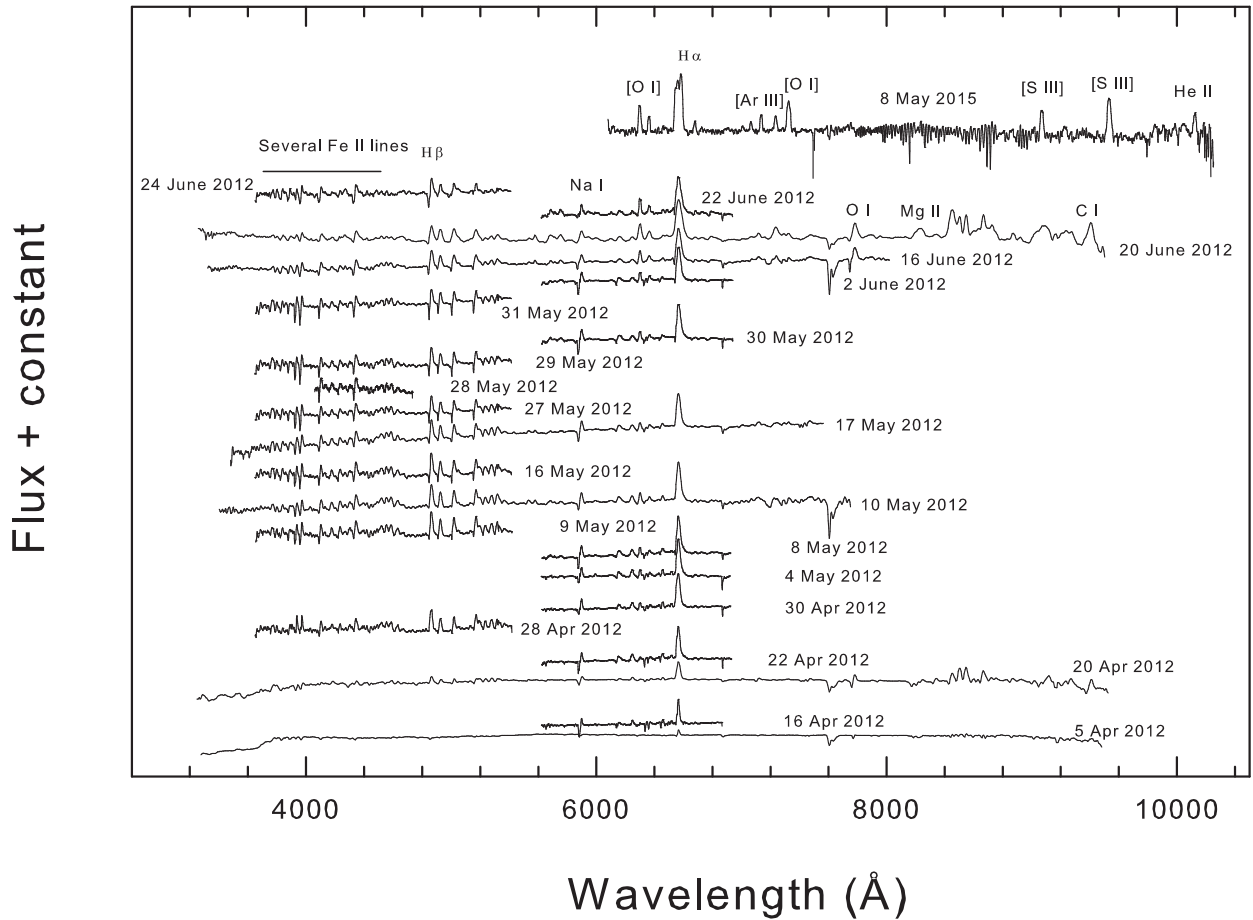}
  \caption{The low-resolution optical spectra of V2676 Oph is presented at different epochs to show the spectroscopic evolution from pre-maximum to post-maximum phase. 
  These observations have been taken from SMARTS and Asiago.}
 \end{figure*}

\section{Acknowledgements}

The research work at Physical Research Laboratory is funded by the Department of Space, Government of India. 
Access to SMARTS is made possible by generous support from the Provost of Stony Brook University, Dennis Assanis. Based in part on observations at Cerro Tololo Inter-American Observatory,
National Optical Astronomy Observatory (NOAO Prop. 2015A-0261, PI F.M. Walter, plus SMARTS time), which is operated by the Association of Universities for Research in Astronomy
(AURA) under a cooperative agreement with the National Science Foundation. 
We thank the referee Prof. A. Evans for his helpful comments as well as Prof. G. C. Anupama (IIA), Dr. U. S. Kamath (IIA) and Dr. B. C. Lee (KASI) for their useful discussions. 
We acknowledge the use of AAVSO (American Association of Variable Star Observers) optical photometric data and 
the spectrophotometric data which was privately communicated by the Asiago ANS (Asiago Novae and Symbiotic stars) Collaboration team.

\begin{table*}
\scriptsize
\caption{Log of the spectroscopic observations of nova V2676 Oph from Asiago 1.22m and SMARTS 1.5m telescopes.}

\begin{tabular}{lcccccc}
\hline \\

Date of&      Days after&Time (s) & Mode  & Resolution & Observatory   \\
observation&         discovery&  &   & &   or \\
(UT) &        &&       &  &Instrument\\

\hline \\
(2012) &  && & &\\
Apr 04  &  10  & 1500 & 13/I &   17 \AA &    SMARTS   \\
Apr 16  &  22& 900 & 47/Ib &   3.1 \AA &     SMARTS   \\
Apr 20   & 26 & 900 & 13/I &   17 \AA  &       SMARTS\\   
Apr 22  &  28 &  900 & 47/Ib  &  3.1 \AA&     SMARTS \\
Apr 28   &  34& 900 & 26/Ia &   4.4 \AA &     SMARTS \\
Apr 30   & 36&900 & 47/Ib   & 3.1 \AA &        SMARTS\\
May 04   &40 &    900 & 47/Ib  &  3.1 \AA&       SMARTS\\
May 08   & 44&900 & 47/Ib   & 3.1 \AA &         SMARTS\\
May 09  &  45& 900  & 26/Ia &   4.4 \AA &      SMARTS \\
May 10 &   46&        900&    --& 2.31 \AA&    ASIAGO \\
May 16   &52&900  &26/Ia  &  4.4 \AA &          SMARTS\\
May 17 &     53 &     1200&    --& 2.31 \AA&    ASIAGO\\
May 27   &  63  &900 & 26/Ia &   4.4 \AA &     SMARTS\\
May 28   &64&1800  &47/IIb   &1.6 \AA &       SMARTS \\
May 29   &   65  &900 & 26/Ia &   4.4 \AA &   SMARTS \\
May 30   & 66 &900 & 47/Ib   & 3.1 \AA &     SMARTS \\
May 31   &  67 &  900 & 26/Ia &   4.4 \AA &  SMARTS \\
Jun 02   & 69& 900 & 47/Ib   & 3.1 \AA &    SMARTS \\
June 16 &     83  &    1800&    --& 2.31 \AA&    ASIAGO\\
Jun 20   &   87 & 900&  13/I &   17 \AA &   SMARTS \\
Jun 22   &89&900  &47/Ib   & 3.1 \AA &    SMARTS \\
Jun 24   &  91 & 900 & 26/Ia &   4.4 \AA &   SMARTS\\
(2015) & & & &&\\
May  08 &  1138&3600 & -- & 3 \AA&   CTIO\\ 
\hline
\end{tabular}
\end{table*}

\begin{table}
\caption[]{The reddening-corrected fluxes (in erg s$^{-1}$ cm$^{-2}$) for prominent emission line corrected for A$_V$ = 2.89 is given in the Table. The complete
table is available in the online version of the paper. }
\begin{tabular}{llcc}
\hline\\
Wavelength & Species  & April 28    &  May 10\\
(\AA) &                &   &\\
\hline \\
3934   & Ca\,{\sc ii}       &1.37E-10 & 1.96E-11\\
3970   & Ca\,{\sc ii} and H$\epsilon$         &1.49E-10 &2.00E-11\\
4023   & He\,{\sc i}         &1.57E-11  &\\
4101   & H$\delta$         &9.89E-11 &2.08E-11\\
4129   &   Fe\,{\sc ii}(27)            &2.52E-11  &\\
4173   & Fe\,{\sc ii}(27) & 3.06E-11&\\
4178   & Fe\,{\sc ii}(28)   & &1.45E-11\\
4233   & Fe\,{\sc ii}(27)   &4.73E-11 &1.42E-11\\
4273   & Fe\,{\sc ii}(27)         &1.09E-11 &5.24E-12\\
4303          &  Fe\,{\sc ii}(27)         & 3.35E-11& 1.71E-11\\
4340   & H$\gamma$    &1.21E-10 &3.08E-11\\
4378   & He\,{\sc i}         &6.52E-12  &\\
4488   & N\,{\sc ii}          & 1.58E-11  & \\
4555 & Fe\,{\sc ii}(37)  &2.81E-11   &8.50E-12\\
4586   & Fe\,{\sc ii}(38)    &5.42E-11  &\\
4634   & N\,{\sc iii}             & 3.77E-11&9.64E-12\\
4861   & H$\beta$                   & 1.16E-10&5.59E-11\\
4924   & Fe\,{\sc ii}(42)                   & 7.96E-11 & 2.43E-11\\
5018   & Fe\,{\sc ii}(42)                  & 7.25E-11   &2.32E-11\\
5169   & Fe\,{\sc ii} + Mg\,{\sc i}             &6.03E-11  &1.86E-11\\
5235   &      Fe\,{\sc ii}(49)               &3.24E-11  &7.83E-12 \\
5276   & Fe\,{\sc ii}(49+48)             & 4.46E-11& 1.04E-11\\
5316   & Fe\,{\sc ii}(49)             & 1.84E-11 &1.48E-11\\
5361   & Fe\,{\sc ii}(48)            & &3.97E-12\\
5528   & Mg\,{\sc i}        & &3.41E-12\\
5577   & [O\,{\sc i}]         &&4.10E-12\\
5676   & N\,{\sc ii}    &&2.83E-12\\
5755 & [N\,{\sc ii}](3)     & &2.53E-12\\
5890-5896 & Na\,{\sc i}                &    &1.31E-11\\
5991 & Fe\,{\sc ii}(46)          &  &1.92E-12\\
6154-6160   & Na\,{\sc i}           &  &7.31E-12\\
6243 & Fe\,{\sc ii} + N\,{\sc ii}       &    &7.63E-12\\
6300 & [O\,{\sc i}]           &  &1.25E-11\\
6364    & [O\,{\sc i}]         &   &3.58E-12\\
6419              & Fe\,{\sc ii}(74)        &   &1.04E-12\\
6456              & Fe\,{\sc ii}          & &3.48E-12 \\
6563              & H$\alpha$          &  &1.89E-10 \\
\hline
\end{tabular}
\label{table4}
\end{table}

\begin{table}
\caption[]{The reddening-corrected fluxes (in erg s$^{-1}$ cm$^{-2}$) for prominent emission line corrected for A$_V$ = 2.89 is given in the Table. 
The complete table is available in the online version of the paper. }
\begin{tabular}{llcccc}
\hline\\
Wavelength & Species &  May 17   & June 16\\
(\AA) & &                     &\\
\hline \\
3934   & Ca\,{\sc ii}      & 4.66E-11 & 3.75E-11\\
3970   & Ca\,{\sc ii} and H$\epsilon$       &4.81E-11  &4.42E-11\\
4101   & H$\delta$        & 3.55E-11 &4.18E-11\\
4178   & Fe\,{\sc ii}(28)  &2.90E-11&1.50E-12\\
4233   & Fe\,{\sc ii}(27)   &2.28E-11 &2.33E-11\\
4340   & H$\gamma$     &4.89E-11 &5.77E-11\\
4586   & Fe\,{\sc ii}(38)  &2.22E-11&1.39E-12\\
4634   & N\,{\sc iii}            &1.50E-11  &1.26E-11\\
4861   & H$\beta$                 & 6.33E-11   &5.19E-11\\
4924   & Fe\,{\sc ii}(42)             & 5.00E-11          &5.17E-11 \\
5018   & Fe\,{\sc ii}(42)             &  4.62E-11     &4.81E-11\\
5169   & Fe\,{\sc ii} + Mg\,{\sc i}         & 8.27E-11       &6.36E-11\\
5235   &      Fe\,{\sc ii}(49)          &   1.91E-11         & 7.38E-12\\
5276   & Fe\,{\sc ii}(49+48)         & 2.35E-11     &1.33E-12 \\
5316   & Fe\,{\sc ii}(49)            & 3.17E-11 &1.94E-12\\
5577   & [O\,{\sc i}]        & 3.26E-12&3.74E-12\\
5890-5896 & Na\,{\sc i}           &  2.96E-11    &1.99E-11\\
6154-6160   & Na\,{\sc i}         & 2.49E-11  &1.25E-11\\
6243 & Fe\,{\sc ii} + N\,{\sc ii}     &  2.02E-11      &1.13E-11\\
6300 & [O\,{\sc i}]         & 1.65E-11&1.49E-11\\
6364    & [O\,{\sc i}]          & 7.53E-12&7.81E-12\\
6456              & Fe\,{\sc ii}          & 1.15E-11          &5.44E-12 \\
6563              & H$\alpha$         & 4.99E-11      & 3.56E-10\\
6678              & He\,{\sc i}         &  3.33E-12    & 3.36E-11\\
7120              & C\,{\sc i} &1.07E-11    &5.29E-12\\
7237              &  [Ar\,{\sc iv}]      &1.79E-11      &1.47E-11\\
7477   & O\,{\sc i}          &1.73E-11   & 8.22E-12\\
\hline
\end{tabular}
\label{table4}
\end{table}

\begin{table}
\caption[]{The reddening-corrected fluxes (in erg s$^{-1}$ cm$^{-2}$) for prominent emission line corrected for A$_V$ = 2.89 is given in the Table. The complete
table is available in the online version of the paper. }
\begin{tabular}{llccc}
\hline\\
Wavelength & Species & June 20 &  2015 May 8 \\
(\AA) & &                 & \\
\hline \\
3934   & Ca\,{\sc ii}      &  2.89E-11  &  \\
3970   & Ca\,{\sc ii} and H$\epsilon$      &  3.43E-11  & \\
4101   & H$\delta$        &5.51E-12  & \\
4178   & Fe\,{\sc ii}(28) &1.65E-12 & \\
4340   & H$\gamma$  &  7.84E-11 & \\
4555 & Fe\,{\sc ii}(37) &6.46E-12 & \\
4663   & Al\,{\sc ii}       & 2.94E-12   & \\
4861   & H$\beta$              &1.45E-11       & \\
4924   & Fe\,{\sc ii}(42)           & 4.71E-11        &  \\
5018   & Fe\,{\sc ii}(42)            & 6.27E-12     & \\
5169   & Fe\,{\sc ii} + Mg\,{\sc i}   & 4.08E-12         & \\
5235   &      Fe\,{\sc ii}(49)         & 3.28E-12      &  \\
5276   & Fe\,{\sc ii}(49+48)        & 6.61E-12       &  \\
5316   & Fe\,{\sc ii}(49)         & 1.35E-11    &  \\
5535   & Fe\,{\sc ii}(55) + N\,{\sc ii} & 2.66E-12     &  \\
5577   & [O\,{\sc i}]      & 9.27E-12  &\\
5755 & [N\,{\sc ii}](3)    &1.13E-11 & \\
5890-5896 & Na\,{\sc i}          &1.88E-11     & \\
5942 & N\,{\sc ii}(28)          & 2.54E-12    & \\
6154-6160   & Na\,{\sc i}       & 9.73E-12   & \\
6243 & Fe\,{\sc ii} + N\,{\sc ii}   &4.80E-12        & 1.89E-13\\
6300 & [O\,{\sc i}]    & 3.69E-11       & 4.84E-12\\
6364    & [O\,{\sc i}]     & 1.54E-11     & 1.59E-12 \\
6563              & H$\alpha$       & 8.20E-11    & 1.76E-10 \\
6678              & He\,{\sc i}      & 2.23E-12  & 8.24E-13 \\
6716-6730              & [S\,{\sc ii}]    &2.64E-12   & 1.84E-13      \\
7002              &        [O\,{\sc i}]  &   &2.23E-13       \\
7065              & He\,{\sc i} &  & 5.32E-13\\
7281   & He\,{\sc i}      &   & 9.04E-14\\
7319-7331              & [O\,{\sc ii}]  &     &5.65E-12     \\
9069              & [S\,{\sc iii}]      &     & 1.57E-12 \\
9531              & [S\,{\sc iii}]      &    &4.21E-12  \\
10126              &  He\,{\sc ii}      &     &8.16E-13 \\
\hline
\end{tabular}
\label{table4}
\end{table}

\end{document}